# Inverse Iron Isotope Effect on the transition temperature of the $(Ba,K)Fe_2As_2$ superconductor

Parasharam M. Shirage[1], Kunihiro Kihou[1], Kiichi Miyazawa[1,2], Chul-Ho Lee[1,3], Hijiri Kito[1,3], Hiroshi Eisaki[1,3], Takashi Yanagisawa[1], Yasumoto Tanaka[1], Akira Iyo[1,2,3]*

[1] National Institute of Advanced Industrial Science and Technology, Tsukuba, Ibaraki 305-8568, Japan

[2] Department of Applied Electronics, Tokyo University of Science, 2641 Yamazaki, Noda, Chiba 278-3510, Japan

[3] JST, Transformative Research-Project on Iron Pnictides (TRIP), 5, Sanbancho, Chiyoda, Tokyo 102-0075, Japan

E-mail: paras-shirage@aist.go.jp, iyo-akira@aist.go.jp

*Corresponding author

**Abstract:**

We report that $(Ba,K)Fe_2As_2$ superconductor (a transition temperature, $T_c \sim 38$ K) has inverse iron isotope coefficient $\alpha_{Fe}$ = -0.18(3) (where $T_c \sim M^{-\alpha_{Fe}}$ and $M$ is the iron isotope mass), i.e. the sample containing the larger iron isotope mass depicts higher $T_c$. Systematic inverse shifts in $T_c$ were clearly observed between the samples using three types of Fe-isotopes ($^{54}Fe$, natural Fe and $^{57}Fe$). This indicates the first evidence of the inverse isotope effect in high-$T_c$ superconductors. This anomalous mass dependence on $T_c$ implies the exotic coupling mechanism in the Fe-based superconductors.

KEYWORDS: isotope effect, high-$T_c$ superconductivity, iron-based superconductor, $(Ba,K)Fe_2As_2$, high-pressure synthesis

PACS numbers: 74.70.Dd, 74.62.Bf

Discovery of the superconductivity with a $T_c$ of 26 K in the Fe-based oxypnictide $LaFeAsO_{1-x}F_x$ has provided a new paradigm for exploring new high-$T_c$ superconductors [1-2]. Although several recent studies suggest that superconductivity in the iron-based system is unconventional and suggesting pairing mechanism is not due to phonon [3-5], the mechanism of superconductivity is currently in dispute. In conventional superconductors, a strong effect of changing ion mass $M$ on $T_c$ implies that lattice vibrations (phonons) play an important role in the microscopic mechanism of superconductivity [6-7]. The isotope exponent $\alpha$ is defined by $T_c \sim M^{-\alpha}$, where $M$ is the isotopic mass. In the classical form of BCS theory [8], $\alpha$ is equal to 0.5. For simple low-$T_c$ metals like Hg, Pb, Sn, and Zn, the isotope coefficient is found experimentally to be close to 0.5. $(Ba,K)BiO_3$ ($T_c \sim$ 30 K), $MgB_2$ ($T_c \sim$ 40 K) and $CaC_6$ ($T_c \sim$ 11 K) show large isotope effect coefficients $\alpha_O \approx 0.4$, $\alpha_B \approx 0.3$ and $\alpha_{Ca} \approx 0.5$, respectively [9-12]. Thus the observation of isotope effect on $T_c$ is crucial in suggesting phonon-mediated paring mechanism in superconductors.

Here we report on the iron isotope effect on $T_c$ for the $(Ba,K)Fe_2As_2$ superconductor ($T_c \approx$ 38 K) [13]. Apparent inverse isotope effect has been observed between the samples containing two different iron isotopes. Possible explanation of the inverse isotope effect is also discussed.

We used three sorts of Fe, $^{57}$Fe enriched powder (ISOFLEX USA, purity 99.99%), $^{54}$Fe enriched powder (ISOFLEX USA, purity 99.99%) and natural Fe powder ($^{n}$Fe) (Showa Chemical Industry, purity 99.9%). The isotopic mass of the $^{57}$Fe, $^{n}$Fe and $^{54}$Fe powders are 57.00, 55.85 and 54.01, respectively.

Polycrystalline samples of $(Ba,K)Fe_2As_2$ were prepared by high-pressure synthesis method [14]. The sample synthesized from a stoichiometric starting composition of $(Ba_{1-x}K_x)Fe_2As_2$ includes a FeAs impurity phase probably because K and Ba exclude from the samples during heat treatment. This difficulty was overcome by using the 50% excess of Ba and K for the starting composition i.e. $(Ba_{0.5}K_{0.5})_{1.5}Fe_2As_2$. Precursors of BaAs, KAs and As powder were mixed homogeneously and divided into two equal weights, then mixed with the two different iron isotopes in order to rule out K concentration difference between the two samples which can be a factor to change $T_c$. The materials were ground by an agate mortar in a glove box. The mixed powders were pressed into pellets. Two pellets were put into a BN crucible as shown in the inset of Fig.1; to rule out difference in synthesis temperature and pressure, which also can affect in a variation in the $T_c$ between the samples. The samples were simultaneously heated at about 900°C under a pressure of about 1 GPa for 1 h. Three combinations of isotopic Fe, $^{n}$Fe and $^{57}$Fe, $^{57}$Fe and $^{54}$Fe, $^{54}$Fe and $^{n}$Fe were tested for sample preparation to ensure systematic behavior of $T_c$. We obtained 7 sets of samples (S1-S7) synthesized in the same conditions except for the iron isotope mass.

Powder X-ray diffraction (XRD) patterns of the samples were measured by using $CuK_a$ radiation. The dc magnetic susceptibility was measured by using a SQUID magnetometer (Quantum Design MPMS) under a magnetic field of 5 Oe. The resistivity of the two isotope containing samples was measured simultaneously by a four-probe method (Quantum Design PPMS).

All the samples are an almost single-phase and difference in lattice parameters between the two samples synthesized simultaneously was small (negligible). The XRD patterns of the sample set S2 synthesized using $^{54}$Fe and $^{57}$Fe isotopic powders, S2($^{54}$Fe) and S2($^{57}$Fe) are represented in Fig. 1 as a typical example. The lattice parameters of the samples S2($^{57}$Fe) and S2($^{54}$Fe) are $a$=3.914(1) Å, $c$=13.310(1) Å and $a$=3.914(1) Å, $c$=13.313(1) Å, respectively. Judging from the $c$-axis parameters, actual K concentration $x$ of the samples in $(Ba_{1-x}K_x)Fe_2As_2$ is about 0.4 [15], where $T_c$ is almost constant against $x$, which is suitable for verification of the isotope effect. We also confirmed the homogeneity of the samples by using scanning electron microscope with energy dispersive x-ray spectroscopy (SEM-EDX) analysis, and we detected the $x = 0.40(2)$ throughout the pellet.

Figures 2 (a)-(c) represent the temperature ($T$) dependence of normalized zero-field cooled (ZFC) susceptibility ($\chi$) for the three type of Fe-isotopic combinations of the samples. The samples demonstrated the $\chi$ at 5 K ranging from -0.0154 to -0.0182 emu/g that corresponds to the superconducting volume fraction of 114-134% without demagnetizing field correction. A clear isotope shift in $T_c$ can be seen in each combination. Surprisingly, the sample containing the larger iron isotope mass depicted higher $T_c$, indicates the inverse isotope effect. $T_{c(\chi)}$ is determined from the definition as shown in the inset of Fig. 2. Isotope shifts in $T_{c(\chi)}$ ($\Delta T_{c(\chi)}$) and $\alpha_{Fe}$ of the sample sets S1-S7 are listed in Table 1.

In Fig. 3 the temperature dependent normalized resistivity of the samples S2($^{57}Fe$) and S2($^{54}Fe$) are shown. It should be noted that the normal state in the normalized resistivity plot is almost overlapped each other except the region of $T_c$, assuring the same quality of the samples. $T_{c(\rho)}$ is determined by the definition as shown in the inset of Fig. 3. This result also clearly shows that sample containing $^{57}Fe$ has higher $T_c$ than $^{54}Fe$. The isotope exponent is estimated to be $\alpha_{Fe} = -0.22(2)$, which is somewhat larger than that measured by magnetic susceptibility method, but comparable.

Fig. 4 shows the iron isotope mass dependent $T_c$ for the sample sets S1-S7. The sample containing the larger iron isotope mass invariably showed higher $T_c$. The results presented here evidently demonstrate that iron isotope effect coefficient ($\alpha_{Fe}$) is negative and highly reproducible. The average $\alpha_{Fe}$ of the 7 sets of samples is -0.18(3). Our result is in contradictory with the earlier report on (Ba,K)$Fe_2As_2$ ($\alpha_{Fe}$ = 0.37) by Liu *et al.* [16]. They synthesized the samples in a quartz tube from a stoichiometric composition of $(Ba_{0.6}K_{0.4})Fe_2As_2$. The contradiction may result from the difference in K concentration of the samples. In the case of cuprate superconductors, a nearly zero oxygen-isotope effect is observed in optimally-doped state, but an oxygen-isotope effect in underdoped state is large [17-22]. It will be necessary to measure the $\alpha_{Fe}$ as a function of K concentration.

The inverse isotope effect in (Ba,K)$Fe_2As_2$ is the first case in high-$T_c$ superconductors [23] strongly suggests that superconducting mechanism of the iron-based system is not explained by conventional BCS theory mediated by phonons, on the other hand, it indicates that phonons play some particular role in this superconductivity. The inverse isotope effect can be explained by assuming a competition between the conventional phonon-mediated superconductivity and the unconventional one (possibly mediated by antiferromagnetic (AF) fluctuation) [24, 25]. When BCS pair hopping between electron Fermi-surface (FS) and hole FS by AF fluctuation dominates the superconductivity, it reverses the sign of the order parameter between two FS's (it is sometimes designated as $s_{+-}$ wave). However, there is also a BCS pair hopping between the two FS's by phonons, which does not reverse the sign ($s_{++}$ wave). When $s_{+-}$ wave is realized, $T_c$ is suppressed with increasing of contribution of phonon, i.e., increasing phonon frequency. It leads an inverse isotope effect.

This tendency can be reproduced by the weak coupling BCS theory when there are several bosons mediating BCS pairs [26]. We take into account two kinds of pair hopping process mediated by AF fluctuation and phonon (neglecting contribution from the intra-FS pair hopping). For simplicity we provide, there are two symmetrical FS's. The $T_c$ can be expressed as follows:

$$k_B T_c = 1.13\hbar\omega_{phonon} \exp\left( -\frac{1}{-\lambda_{ph} - \frac{1}{\ln\left(\frac{\omega_{AF}}{\omega_{phonon}}\right) - \frac{1}{\lambda_{AF}}}} \right).$$

$\lambda_{AF}, \omega_{AF}, \lambda_{ph}, \omega_{phonon}$ are dimensionless coupling constants and cut-off frequencies of the AF fluctuation and phonon, respectively. We provide $\omega_{AF} > \omega_{phonon}$, but it is not crucial for the explanation of the inverse isotope effect. Four values are taken as positive values and sign of the pair interaction directly reflect the sign in the formula. This formula can be reduced in the case of $\lambda_{AF} >> \lambda_{ph}$ as follows:

$$k_B T_c \approx 1.13\hbar\omega_{AF} \exp\left( -\frac{1}{\lambda_{AF}} - \lambda_{ph}\left\{ \ln\left(\frac{\omega_{AF}}{\omega_{phonon}}\right) - \frac{1}{\lambda_{AF}} \right\}^2 \right).$$

When cut-off frequency of the AF fluctuation and phonon is comparable, the inverse isotope effect emerges. It is the consequence of a competition between the AF fluctuation and phonon for inter-band BCS pair hopping process.

$\lambda_{ph} / \lambda^*_{AF} = 0.14$ reproduces the inverse isotope shift of $\alpha_{Fe}$ = -0.18,

where $\lambda^*_{AF} = \dfrac{\lambda_{AF}}{1 - \lambda_{AF} \ln\left(\omega_{AF} / \omega_{phonon}\right)}$.

Providing $\omega_{phonon}$ = 350 K, we deduce $\lambda_{ph}$ = 0.07, $\lambda^*_{AF}$ =0.50 gives $T_c$= 38 K as reasonable values. (We note that the intra-band pair hopping mediated by the phonon cooperates with both inter-band processes.)

In conclusion, a significant inverse iron isotope effect ($\alpha_{Fe}$ = -0.18(3)) was observed in (Ba,K)$Fe_2As_2$ superconductor. Results obtained here suspect that phonon plays some particular role and implies the exotic coupling mechanism in this superconductivity. To explain this anomaly will be a key to understand the mechanism of the Fe-based superconductors, may lead to pioneering the basics in the field of high-$T_c$ superconductivity.

**Acknowledgements**: This work was supported by Grant-in-Aid for, Specially promoted Research (20001004) from The Ministry of Education, Culture, Sports, Science and Technology (MEXT) and JST, Transformative Research-Project on Iron Pnictides (TRIP).

Correspondence and requests for materials should be addressed to A. I. (iyo-akira@aist.go.jp) or P.M.S.(paras-shirage@aist.go.jp).

Table 1: Istope shift in $T_{c(\chi)}$ ($\Delta T_{c(\chi)}$) and $\alpha_{\mathrm{Fe}}$. $T_{c(\chi)}$ are determined form the magnetic susceptibility measurements of (Ba,K)$Fe_2As_2$ samples containing $^{54}$Fe, $^{n}$Fe and $^{57}$Fe isotopes. The average $\alpha_{Fe}$ is -0.18(3).

| | $T_{c(\chi)}$ (K) | | | $\Delta T_{c(\chi)}$ | $\alpha_{\mathrm{Fe}}$ |
|---|---|---|---|---|---|
| | $^{54}$Fe | $^{n}$Fe | $^{57}$Fe | | |
| S1 | | 37.79(1) | 37.91(1) | -0.12(2 ) | -0.16(3) |
| S2 | 37.56(1) | | 37.82(2) | -0.26(3 ) | -0.13(1) |
| S3 | 37.51(1) | 37.76(1) | | -0.25(2 ) | -0.20(1) |
| S4 | 37.54(1) | 37.79(2) | | -0.25(2 ) | -0.20(2) |
| S5 | 37.32(1) | | 37.75(1) | -0.43( 2) | -0.21(1) |
| S6 | 37.42(1) | | 37.77(1) | -0.35(2 ) | -0.17(1) |
| S7 | 37.39(1) | | 37.76(1) | -0.37( 2) | -0.18(1) |

**Figure captions**

Fig. 1 X-ray diffraction patterns of the samples S2($^{57}$Fe) and S2($^{54}$Fe) synthesized simultaneously in a BN crucible by high-pressure synthesis technique. The inset shows the sample cell assembly for the high-pressure synthesis method.

Figs. 2 (a)-(c) Temperature ($T$) dependence of susceptibility ($\chi$) of the three sample sets using three different combinations of the isotopic Fe (a) S1($^{n}$Fe) and S1($^{57}$Fe), (b) S2($^{57}$Fe) and S2($^{54}$Fe), (c) S3($^{54}$Fe) and S3($^{n}$Fe). The $\chi$ is normalized at the values of ZFC (zero-field cooled) curve at 5K.

Fig. 3 Temperature ($T$) dependence of resistivity ($\rho$) of the sample S2($^{57}$Fe) and S2($^{54}$Fe). Fe isotope exponent is estimated to be $\alpha_{Fe}$ = -0.22(2).

Fig. 4 The iron isotope mass dependence of $T_{c(\chi)}$ for the 7 sets of the samples S1-S7. Fe isotope exponent $\alpha_{Fe}$ is -0.18(3).

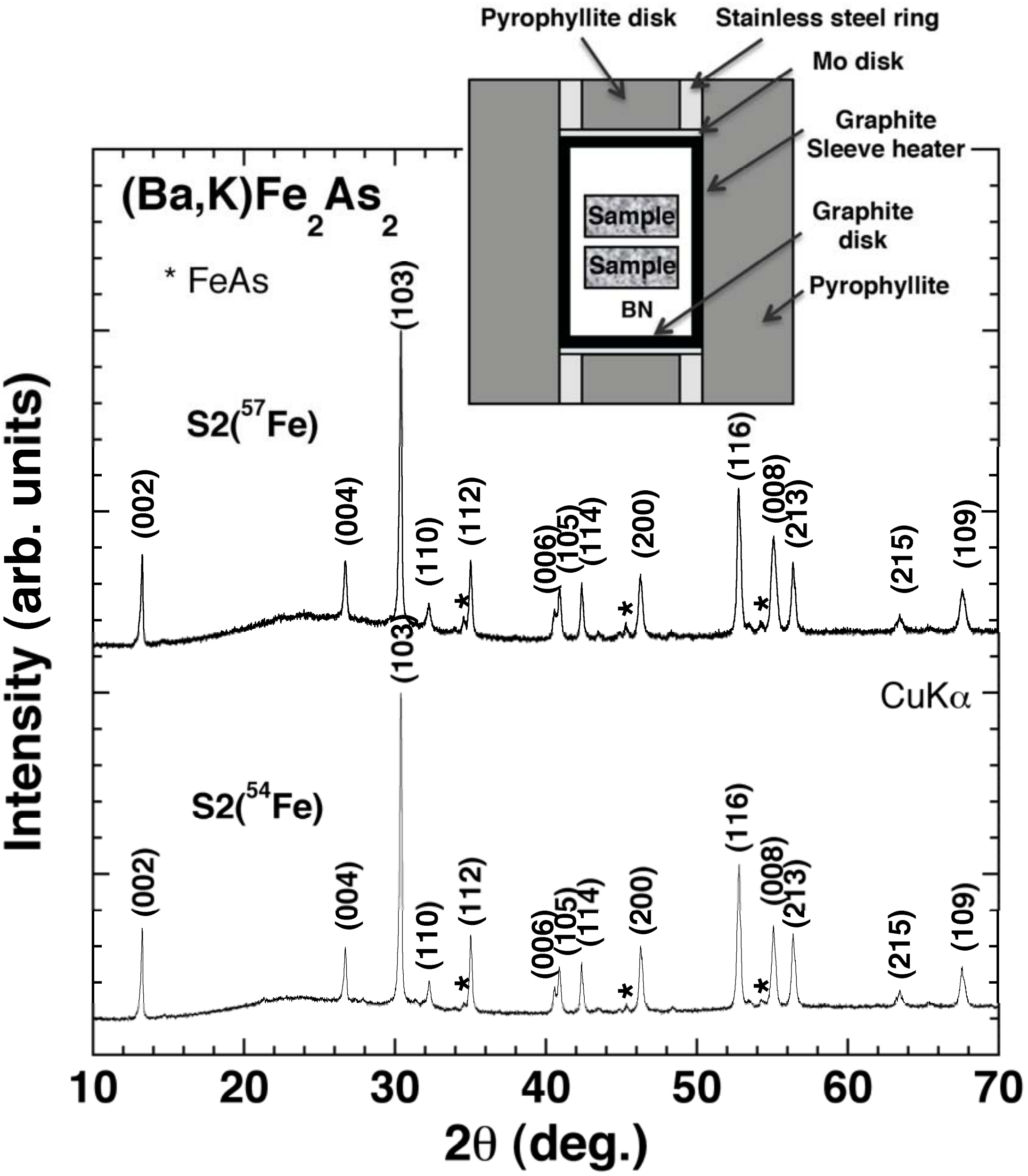

**Shirage *et al.* Figure 1**

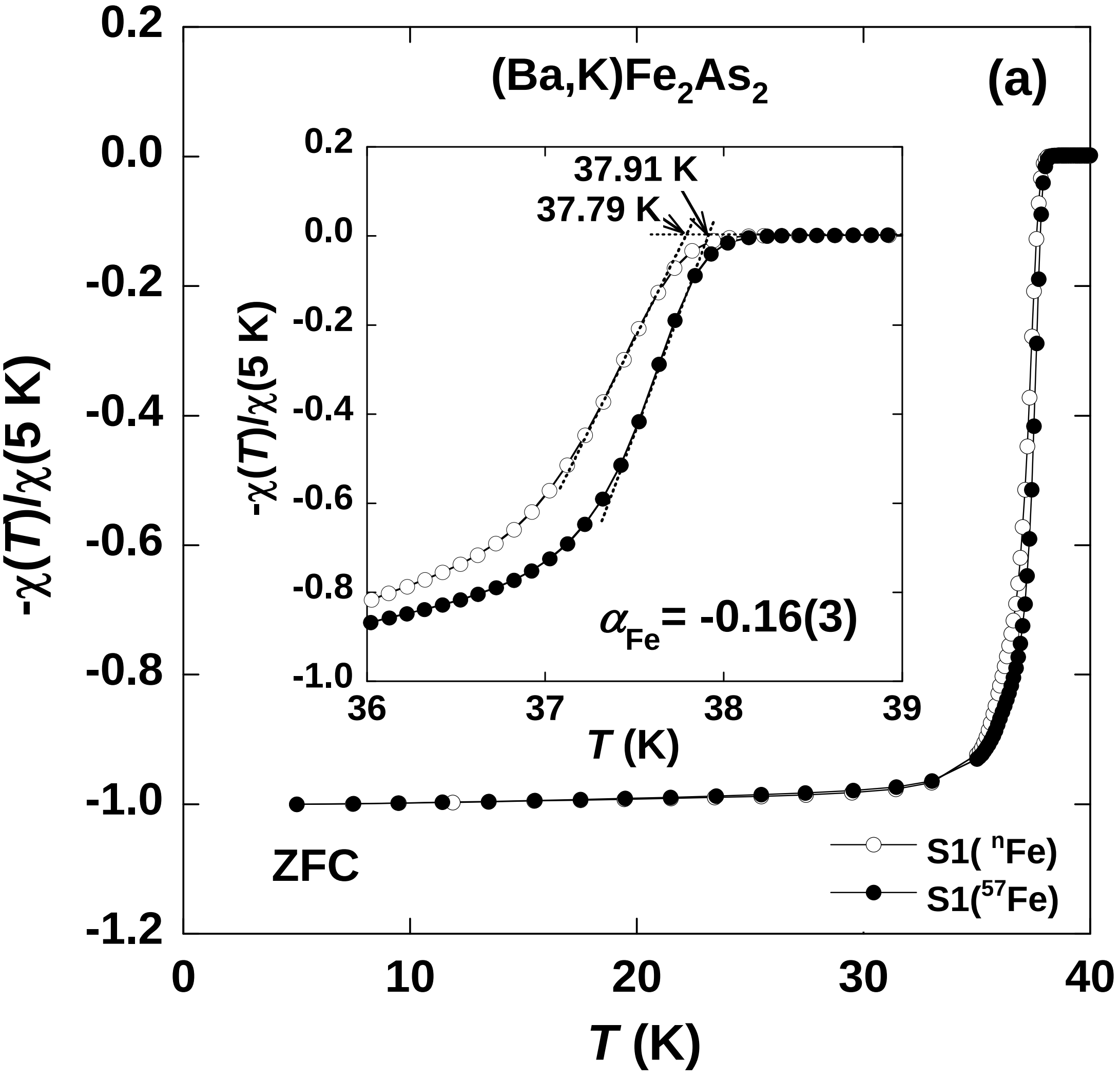

Shirage *et al.* Figure 2 (a)

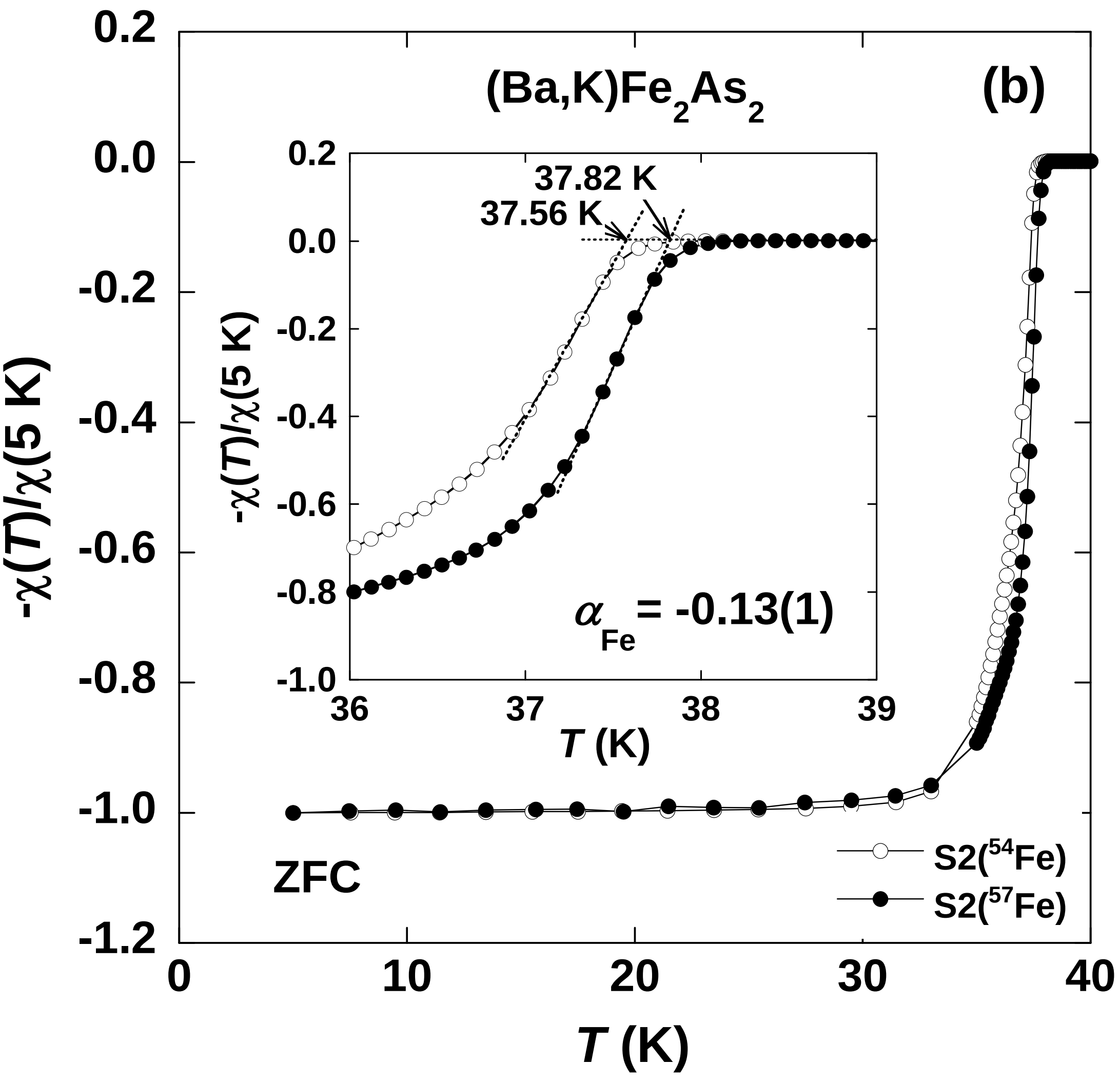

Shirage *et al.* Figure 2 (b)

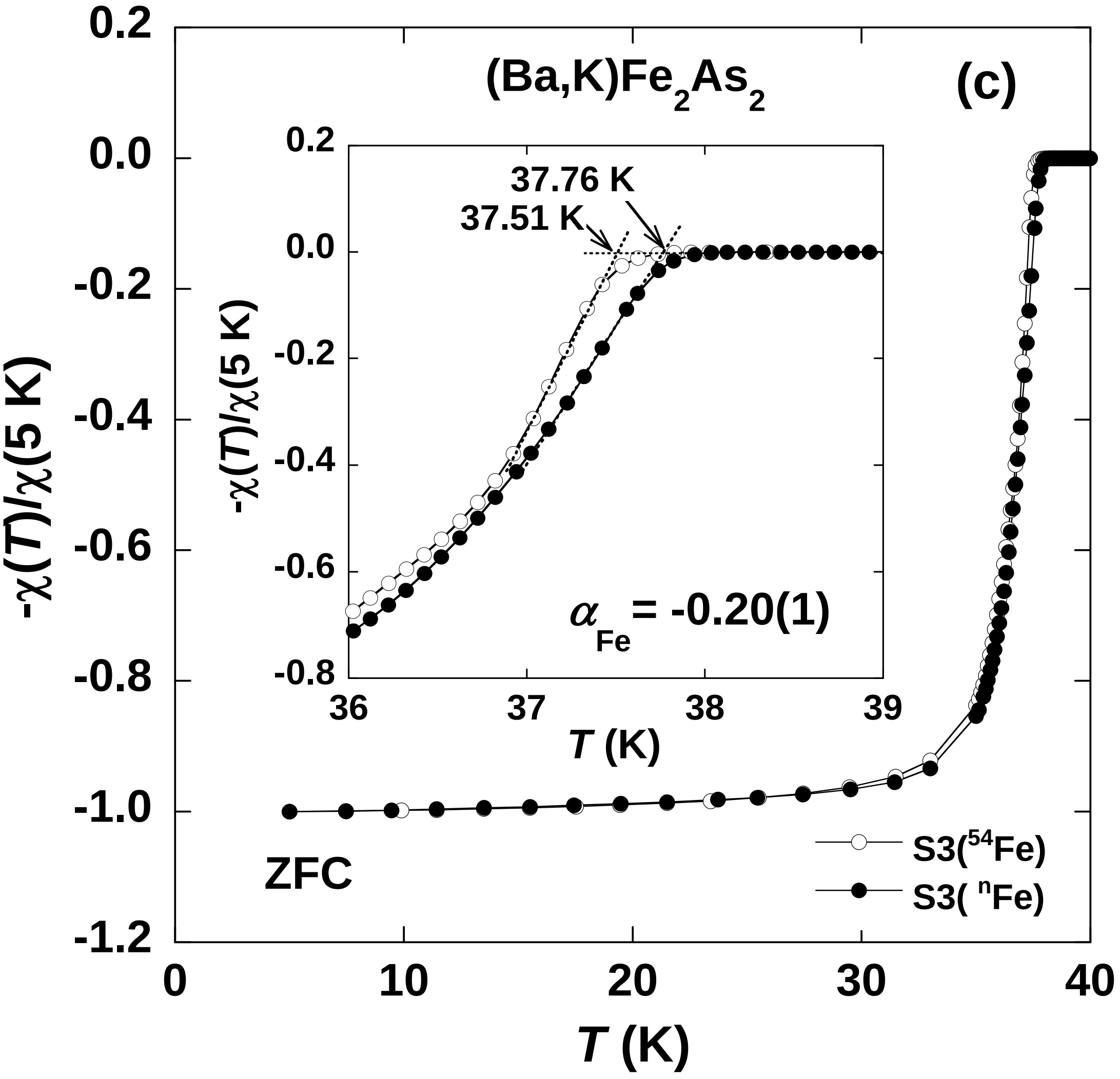

**Shirage *et al.* Figure 2 (c)**

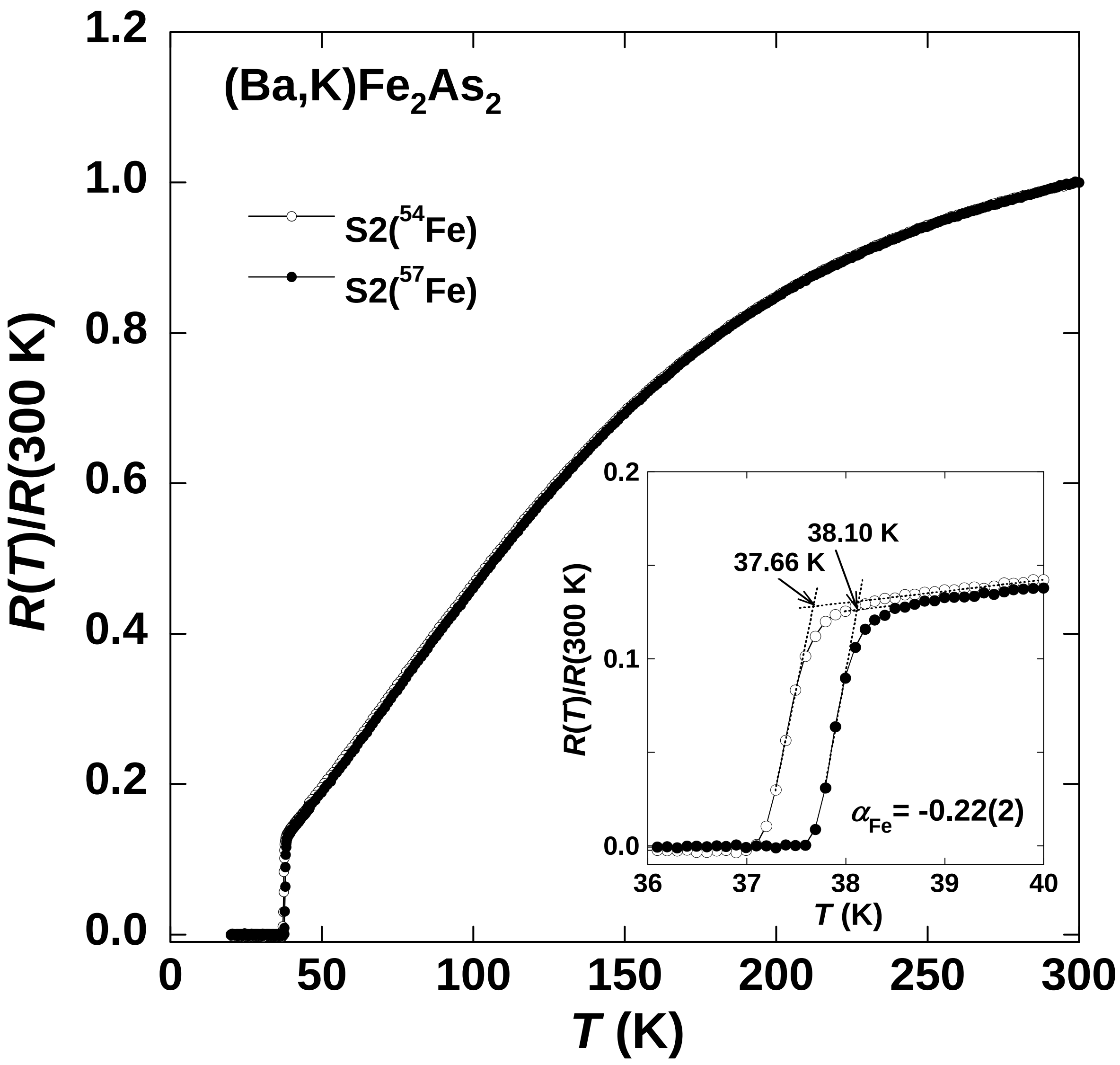

**Shirage *et al.* Figure 3**

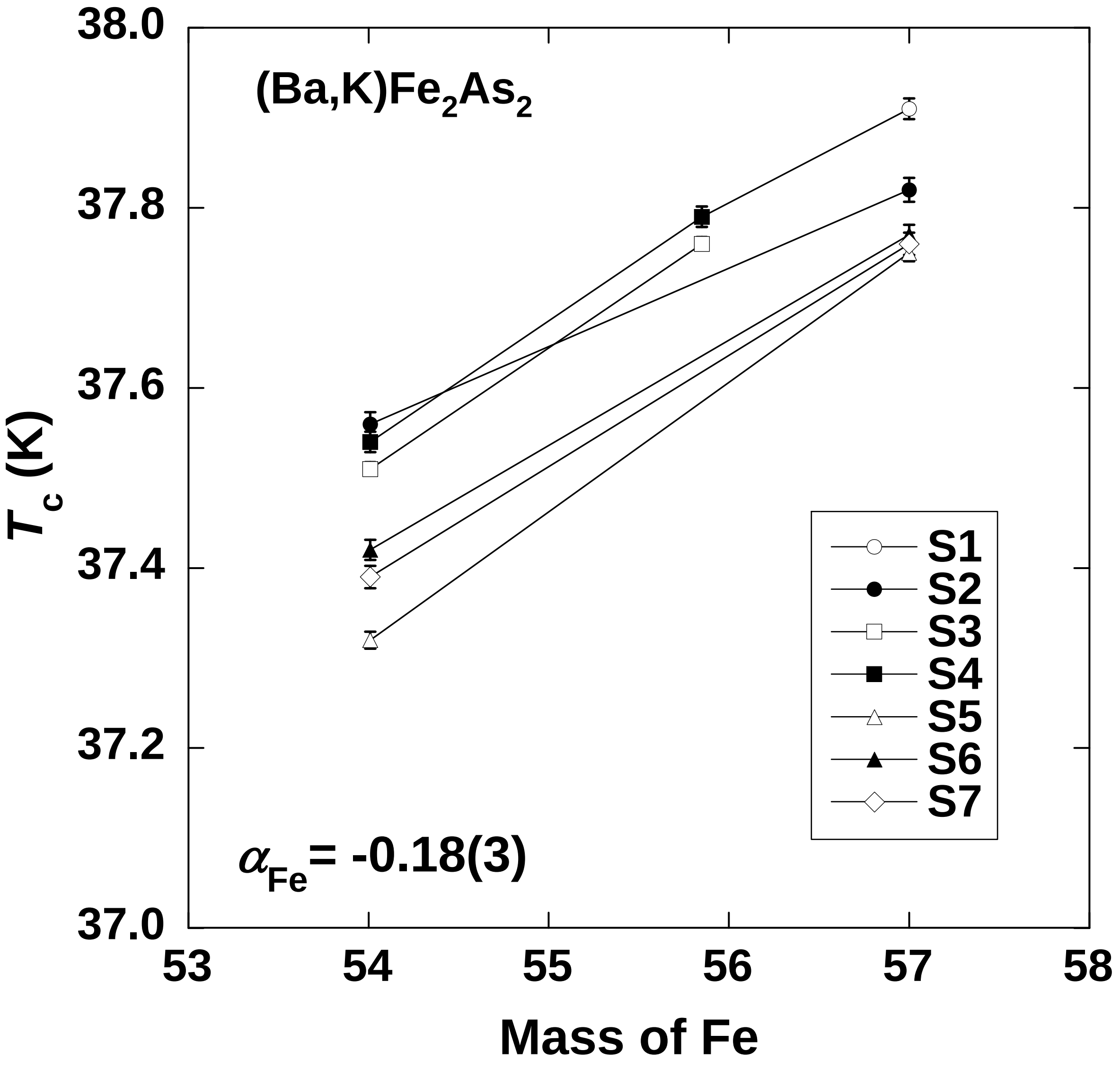

**Shirage *et al.* Figure 4**